\documentclass[10pt]{article}
\pdfoutput=1

\usepackage{amsmath}
\usepackage{amssymb}

\usepackage{graphicx}

\usepackage{cite}

\usepackage{color}
\definecolor{purple}{RGB}{147, 112, 255}

\usepackage{epstopdf}
\usepackage{amsthm}
\usepackage{subfigure}
\usepackage{units}
\usepackage{enumitem}

\usepackage[labelfont=bf,labelsep=period,justification=raggedright]{caption}

\makeatletter
\renewcommand{\@biblabel}[1]{\quad#1.}
\makeatother

\date{}

\begin{document}

\begin{flushleft}
{\Large
\textbf{Sparse signals for the control of human movements using the infinity norm}
}
\\
\vspace{.1cm}
Geoffrey George Gamble$^{1}$,
Mehrdad Yazdani$^{2}$

\vspace{.4cm}
\bf{1} Geoffrey George Gamble $\ast$, Department of Computer Science and Engineering, University of California at San Diego, La Jolla, California, United States of America
\\
\vspace{.2cm}
\bf{2} Mehrdad Yazdani, Qualcomm Institute, University of California at San Diego, La Jolla, California, United States of America
\\
\vspace{.4cm}
$\ast$ E-mail: ggamble@cs.ucsd.edu
\end{flushleft}

\newpage

\section{Abstract}
  Optimal control models have been successful tools in describing many aspects of human movements. While these models have a sound theoretical foundation, the interpretation of such models with regard to the neuronal implementation of the human motor system, or how robotic systems might be implemented to mimic human movement, is not clear. One of the most important aspects of optimal control policies is the notion of \emph{cost}...This body of mathematics seeks to minimize some notion of \emph{cost}, while meeting certain goals. We offer a mathematical method to transform the current methodologies found in the literature from their traditional form by changing the norm by which cost is assessed. In doing so we show how sparsity can be introduced into current optimal control approaches that use continuous control signals. We assess cost using the $L_{\infty}$ norm. This influences the optimization process to produce optimal signals which can be represented by a small amount of Dirac delta functions, as opposed to the traditional continuous control signals. In recent years sparsity has played an important role in theoretical neuroscience for information processing (such as vision). Typically, sparsity is imposed by introducing a cardinality constraint or penalty (measured or approximated by the one-norm). In this work, however, to obtain sparse control signals, the $L_{\infty}$ norm is used as a penalty on the control signal, which is then encoded with Dirac delta functions. We show that, for a basic physical system, a point mass can be moved between two points in a way that resembles human fast reaching movements. Despite the sparse nature of the control signal, the movements that result are continuous and smooth. These control signals are simpler than their non-sparse counterparts, yet yield comparable if not better results when applied towards modeling human fast reaching movements. In addition, such sparse control signals, consisting of Dirac delta functions have a neuronal interpretation as a sequence of spikes, giving this approach a biological interpretation. Actual neuronal implementations are clearly more complex, as they may consist of large numbers of neurons. However, this work shows, in principle, that sparsely encoded control signals are a plausible implementation for the control of fast reaching movements. The presented method could easily be scaled up to arbitrarily large numbers of duplicates, thus representing larger numbers of spikes. We show how leading techniques for modeling human movements can easily be adjusted in order to introduce sparsity, and thus the biological interpretation and the simplified information content of the control signal.

\section{Introduction}
\label{sec:Introduction}
Optimal control theory has provided a great deal of insight with regard to developing mathematical models that describe human movements (for example, \cite{Flash, KawatoMinTorqueChange1989, Hoff93, Harris1998, Liu2007}). These works, amongst many others, have shown that humans move with strategies that can be described/driven by various control signals\footnote{Control signals are to be described in more depth later, but for now it is a signal that controls elements of a system.} and related cost functions to model movement. However, as \cite{BayesianBrain} points out, while the development of optimal control models has given mathematical insights into the properties of human movements, and perhaps the costs that forged our motor system via evolution, the connection to the neuronal implementation of the motor system is not clear. In contrast to these models, we show that a novel penalty on a control signal results in signals which can be represented more simply, and that have more plausible biological interpretations, while maintaining the ability to model human movements accurately.

To demonstrate the utility of sparse optimal control signals for human movements, we will compare two versions of a class of problems called ``minimum effort" control problems, which attempt to minimize the ``size" or ``effort" of the control signal when modeling human movements, as explained in \cite{neustadt1962minimumEffortDefinition}. In the first, and more traditional version, the effort to be minimized is defined as the $L_{2}$ norm of the control signal over some time course within which a movement is completed. The goal is to minimize that signal. One of the first, and most famous of this family of models is the ``minimum jerk" control policy, proposed originally by Flash and Hogan \cite{Flash}. The Flash and Hogan control strategy models human reaching movements, and uses jerk (the third derivative of position) as a control signal, and minimizes that signal to the extent allowed by a well defined reaching task. Intuitively, most are more familiar with thinking of acceleration as a control signal (e.g. pressing the accelerator in a car controls the speed). In the Flash and Hogan case, jerk is the derivative of acceleration, so it ``controls" the acceleration in the same way acceleration ``controls" velocity. Because jerk is the minimized control signal, we refer to jerk as the ``effort term". Because they attempt to minimize their control signal, this as a minimum effort problem, and is referred to as ``minimum jerk".

Work since Flash and Hogan has considered different cost functions as the effort term as defined above, such as minimizing torque over the course of a movement \cite{lim2005movementMinTorque}, or minimizing torque change (derivative of torque) over the movement  \cite{KawatoMinTorqueChange1989, MinTorqueChange}. Other works have added more terms to the a cost function but have maintained an effort term. Additions to the effort term include end-point stability (how much adjustment is needed once the target area of a reaching movement is breached) or end-point accuracy (how close to the target when the reaching movement ends) \cite{Liu2007}. Such extensions, however, have not lead to any insights into the neuronal implementation of control signals in the CNS, nor do they simplify the nature of the control signal.

We will show that using the $L_{\infty}$ norm instead of the $L_{2}$ norm for measuring and penalizing the ``effort" of the control signal results in signals that can be encoded sparsely via Dirac delta functions. There exists a family of models where this technique is applicable, specifically, because they all employ an ``effort" term. Practically, the sparsification of signals generated by this family of optimal control models might be useful in a robotic system in order to achieve human like-movement. Due to the simplicity (sparsity) of the resulting signal, implementations of human-like robotic control may be easier to comprehend and construct.

Models of reaching movements leading to this work were considered in \cite{Ben-Itzhak, Yazdani2012, KarnielMinAccelerationWithConstraintsOfCenterOfMass}. These works referred to their signals as ``bang-bang" or ``intermittent" (see \cite{KarnielOpenQuestionsMotorControl,GawthropIntermittentControlCompTheory} for more on intermittent control). However, the control signals were not sparse, rather, they were \emph{square wave continuous}. Other types of motor control such as standing and keeping balance have been modeled via intermittent control, notably, two such models are compared in \cite{gawthrop2014intermittent}. Here, we demonstrate a mathematical relationship that can convert non-sparsity to sparsity with regard to the control signal. This relationship is related to the metric used to measure the control signal (i.e. how is the effort measured?), but more importantly a sparse encoding of the signal via Dirac delta functions. We also show that sparse optimal control signals model real human arm movements with high accuracy, thus supporting sparse optimal control signals as a plausible control strategy used by a human's biological system. We emphasize that, using sparse signals in the cases shown here has no downside in terms of model performance, but has the benefits of a simpler encoding of the control signal, a biological interpretation in terms of neural spike timing, and potentially, a simpler control strategy which may be useful in robotics...All of these qualities are absent from non-sparse signals.

\subsection{Optimal Control Overview}
\label{subsec:OptimalControlOverview}
In this section we give an overview of optimal control theory and highlight two optimal control problems: the minimum-time and the minimum-effort control problems. Our overview is meant as means to establish common notation and terminology. For a more in depth overview, see \cite{Kirk, BayesianBrain}. Optimal control theory is an application of optimization theory to the control of a dynamic system. In optimization theory, we seek to find an element in a domain that minimizes (or maximizes) a criterion (also referred to as an objective), while satisfying a constraint set. When the elements in the intersection of the domain and constraint sets are functions, the criterion is typically referred to as an objective functional or a cost functional, whereas when the elements are points in a vector space, the criterion is referred to as an objective function or a cost function. In optimal control, we seek an optimal controller (typically a function of time if the system is continuous or a vector if the system is discrete) that has certain constraints (for example, the controller is limited by a specified amount of power or resources) that minimizes a criterion.

We describe dynamic systems as a set of first-order differential equations:
\begin{equation}\label{eq:dynamic}
\dot{\textbf{x}}(t) = \textbf{a}(\textbf{x}(t), \textbf{u}(t), t).
\end{equation}
$\textbf{x}(t)$ is referred to as the ``state" of the system, $\textbf{u}(t)$ is the controller of the system, and $\textbf{a}(\cdot, \cdot, t)$ is, in general, a non-linear time dependent function describing the dynamics of the state as determined by the state and controller at time $t$. We assume that the initial state $\textbf{x}(t_{0})$ and initial time is known. Often the dynamic system is assumed to be linear time-invariant (LTI) and can be expressed as
\begin{equation}\label{eq:LTI}
\dot{\textbf{x}}(t) = \textbf{A}\textbf{x}(t) +\textbf{B}\textbf{u}(t).
\end{equation}

Given the dynamic system of equation \ref{eq:dynamic} and an initial state $\textbf{x}(t_{0})$, we seek a control signal $\textbf{u}(t)$ to transfer the system to a desired state in a finite time. In practice, the control signal $\textbf{u}(t)$ is not unconstrained, but rather bounded by the available resources (such as fuel, energy, or supply). In optimal control theory, we seek an optimal control signal $\textbf{u}^{*}(t)$ that, in addition to transferring the system to a desired state, also minimizes a cost functional $\textbf{J}(\textbf{u}(t))$. The cost functional is application dependent and the optimal solution $\textbf{u}^{*}(t)$ depends on what we consider to be ``cost". For example, in the cost functional we may penalize large control signals or penalize deviations from a desired trajectory. Subsequently we will discuss two important cost functionals.

\subsubsection{Minimum-Time Control}
\label{subsec:MinTimeControl}
In the minimum-time control problem, the objective is to transfer a system to a final state with a constrained control signal as quickly as possibly. Thus, the cost functional penalizes the total time it takes to transfer the initial state to a final state and can be expressed as
\begin{equation}\label{eq:minTime}
\textbf{J}(\textbf{u}(t)) = t_{f} - t_{0}
\end{equation}
where the initial state $\textbf{x}(t_{0})$, initial time $t_{0}$, and final state $\textbf{x}(t_{f})$ are known, while $t_{f}$ is unknown, and the system dynamics are described by equation \ref{eq:dynamic}. We furthermore constrain the control signals to be bounded
\begin{equation}\label{eq:bound}
|\textbf{u}(t)| \leq B.
\end{equation}

We now show the solution to the minimum-time control problem for an LTI system as described by equation \ref{eq:LTI}. We consider $\textbf{A}$ and $\textbf{B}$ to be constant $n \times n$ and $n \times m$ matrices respectively. Thus the minimum-time control problem can be expressed as

\begin{equation}\label{eq:minimumTimeSimple}
\begin{aligned}
& \underset{\textbf{u}(t), T}{\text{minimize}}
& & T   \\
& \text{subject to} & & \dot{\textbf{x}}_{n}(t) = \textbf{A} \textbf{x}_{n}(t) + \textbf{B}\textbf{u}(t) \\
& & & \textbf{x}_{n}(0) = \textbf{x}_{i} \\
& & & \textbf{x}_{n}(T) = \textbf{x}_{f} \\
& & & |\textbf{u}(t)| \leq B
 \end{aligned}
\end{equation}
where we have defined $T \equiv t_{f} - t_{0}$ and assumed $t_{0} = 0$. This special case has been solved by Pontryagin and colleagues. Their conclusions lead to several important points upon which this work builds, as they guarantee a control signal which switches a finite number of times between two possible values. These points are summarized below (see \cite{Pontryagin} for more details).
\begin{enumerate}[label=\textbf{P.\arabic*},ref = P.\arabic*]
  \item For a given LTI system, there is one, and only one optimal signal to drive the system from an initial state to a desired state.
  \item \label{enum:controlSwitchExtremes} Because the goal of a minimum-time problem is to move the system to the desired state in the least amount of time, a control signal representing a dynamic variable (such as acceleration) is always at one of two extremes, $+B$ or $-B$. These extremes are defined in the constraints of equation \ref{eq:minimumTimeSimple}. Intuitively, if you want to get from point $X$ to point $Y$ as fast as possible, you would change from zero acceleration to maximum positive acceleration to speed up initially, and then to maximum deceleration to slow down to reach point $Y$, and then to zero acceleration to maintain your starting position. The two extreme values are the only values that yield to a minimum time (optimal) result.
  \item \label{enum:controlSwitchesLimited} The control signal will switch between these two extremes at most $n+1$ times, where $n$ is the derivative of position we choose to be the control signal. For example, for velocity, acceleration, and jerk, $n$ is $1$, $2$ and $3$ respectively, and thus has a maximum of $2$, $3$, and $4$ (respectively) switches between the extremes of the control signal.

\end{enumerate}

%

This type of control signal is sometimes referred to as a ``bang-bang" control signal since the signal switches between the lower bound and upper bound of the inequality constraint of equation \ref{eq:bound} (see \cite{Kirk} for more). It can also be regarded as a ``sparse" control signal since the number of changes in the signal's values is small. In other words, the changes in the control signal can be described by a bounded number of switches between the lower and upper bound. These switches can be encoded by a series of Dirac delta functions, which resemble neural bursts or spikes.


\subsubsection{Minimum-Effort Control} \label{subsec:miniumEffortControl}
In the minimum-effort control problem, the objective is to transfer a system from an initial state to a final state with a control signal that is as ``small" as possible (hence, minimum ``effort"). Typically, the ``size" of a control signal is measured with a penalty function. In this work we consider the $L_{p}$ norm penalty function and can express the cost functional as
\begin{equation}\label{eq:minLP}
\textbf{J}(\textbf{u}(t)) = \left(\int u(t)^{p}dt\right)^{1/p} 
\end{equation} 
which denotes the $L_{p}$ norm (and is typically the $L_{2}$ norm), and the system dynamics are described by equation \ref{eq:dynamic}. We can also have additional constraints in the minimum-effort control problem, and just as in the minimum-time control problem, there can be many variations by introducing additional constraints or additional costs to the objective. For example, a simple extension would be to consider a control problem where the cost functional trades-off between ``effort" and the transfer time and can be expressed as a combination of equations \ref{eq:minTime} and \ref{eq:minLP}
\begin{equation}\label{eq:tradeoff}
\textbf{J}(\textbf{u}(t)) = \gamma\left(\int u(t)^{p}dt\right)^{1/p} + t_{f} - t_{0} 
\end{equation}
where $\gamma \geq 0$ is a trade-off parameter between ``effort" and the state transfer time and can be varied depending on the application.

As an example of a minimum-effort problem, Flash and Hogan considered the following minimum-effort control problem which introduced constraints on initial and final state in order to describe human movements:
\begin{equation}\label{eq:FandH}
\begin{aligned}
& \underset{u(t)}{\text{minimize}}
& & \left(\int u(t)^{2}dt\right)^{1/2} \\ 
& \text{subject to} & & \dot{\textbf{x}}(t) = A \textbf{x}(t) + Bu(t) \\
& & & \textbf{x}(0) = \textbf{x}_{i} \\
& & & \textbf{x}(T) = \textbf{x}_{f}
 \end{aligned}
\end{equation}
where $\textbf{x}(t) = \begin{bmatrix} x(t) & \dot{x}(t) & \ddot{x}(t) \end{bmatrix}^\top$ is the state vector, $\textbf{x}_{i}$ and $\textbf{x}_{f}$ are the initial and final boundary conditions, and $T$ is the duration of the movement (with movement starting at time $t = 0$). Flash and Hogan used a jerk control signal ($u(t) = \dddot{x}(t)$), and furthermore used a third-order integrator model for the linear time-invariant dynamic equation parameters:
\begin{equation}
\textbf{A} = \begin{bmatrix}
0 & 1 & 0\\
0 & 0 & 1\\
0 & 0 & 0
\end{bmatrix}
\textnormal{ and } \textbf{B} = \begin{bmatrix}
0 \\
0\\
1
\end{bmatrix}.
\end{equation}
This simple model yields trajectories that are remarkably similar to those of humans. Naturally, simple extensions of this optimization problem can yield results that are even more realistic and many researches have begun exploring these extensions. For example, \cite{Harris1998} has noted that when humans make movements to a target, the target that is reached is a not a specific point, but rather a distribution of points. Hence, in their optimization procedure they relaxed the constraints of equation \ref{eq:FandH}, which specify an exact final state.

\subsection{Sparse optimal control policies for straight point-to-point trajectories}
\label{sec:SparseControlPointToPoint}
\begin{figure}
  \centering
  \subfigure[An example of the types of sparsely driven signals generated by minimizing the control signal as measured by the $L_{\infty}$ norm.]{\label{fig:pulse}\includegraphics[width=0.4\textwidth]{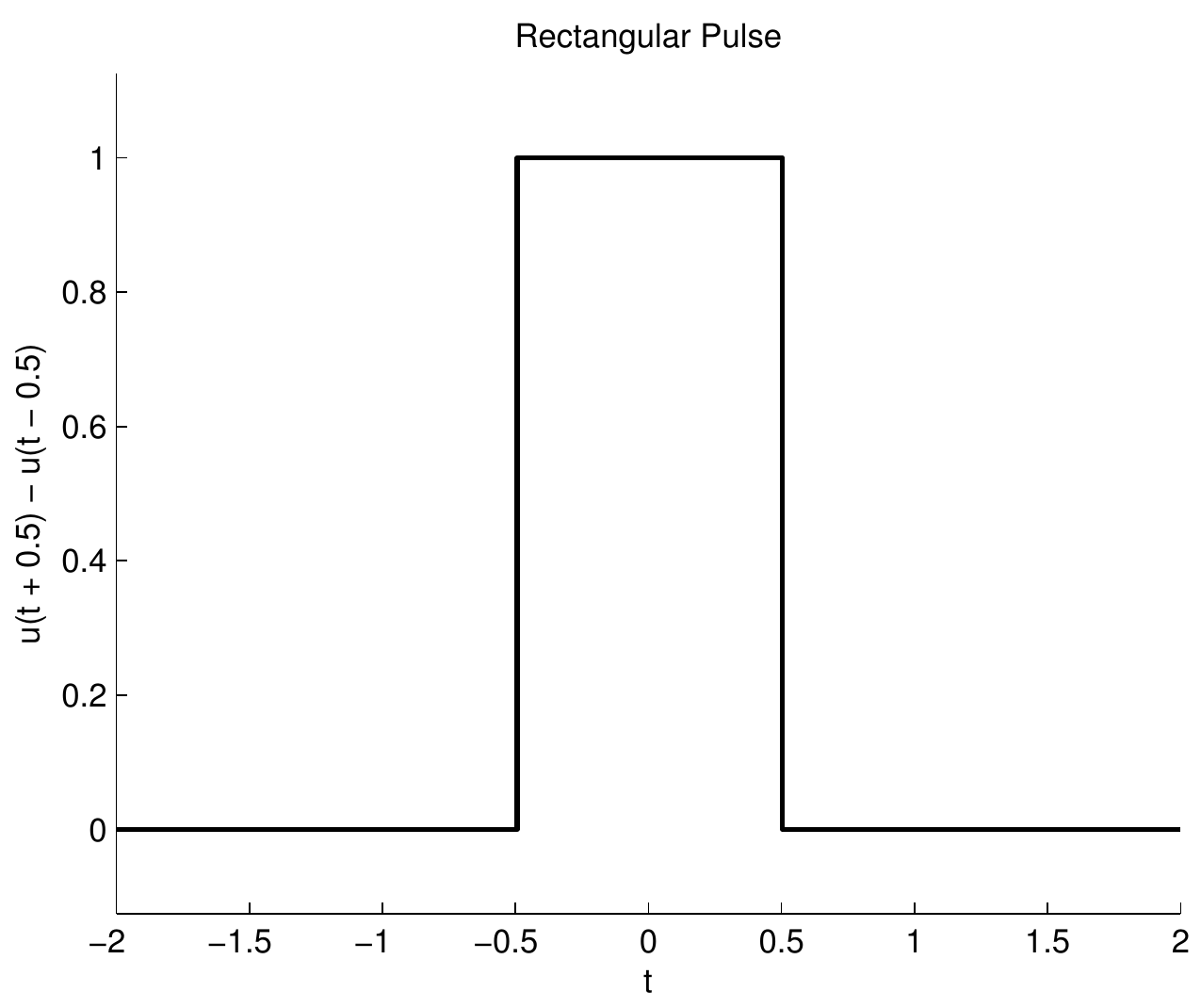}}
  \subfigure[Rate of change of sparse signal shown in Figure \ref{fig:pulse}. These sparse impulse functions can be thought of as driving (or encoding) the control signal in figure \ref{fig:pulse}, along with initial conditions, they are all that is needed to encode the control signal in \ref{fig:pulse}.]{\label{fig:spikedUp}\includegraphics[width=0.4\textwidth]{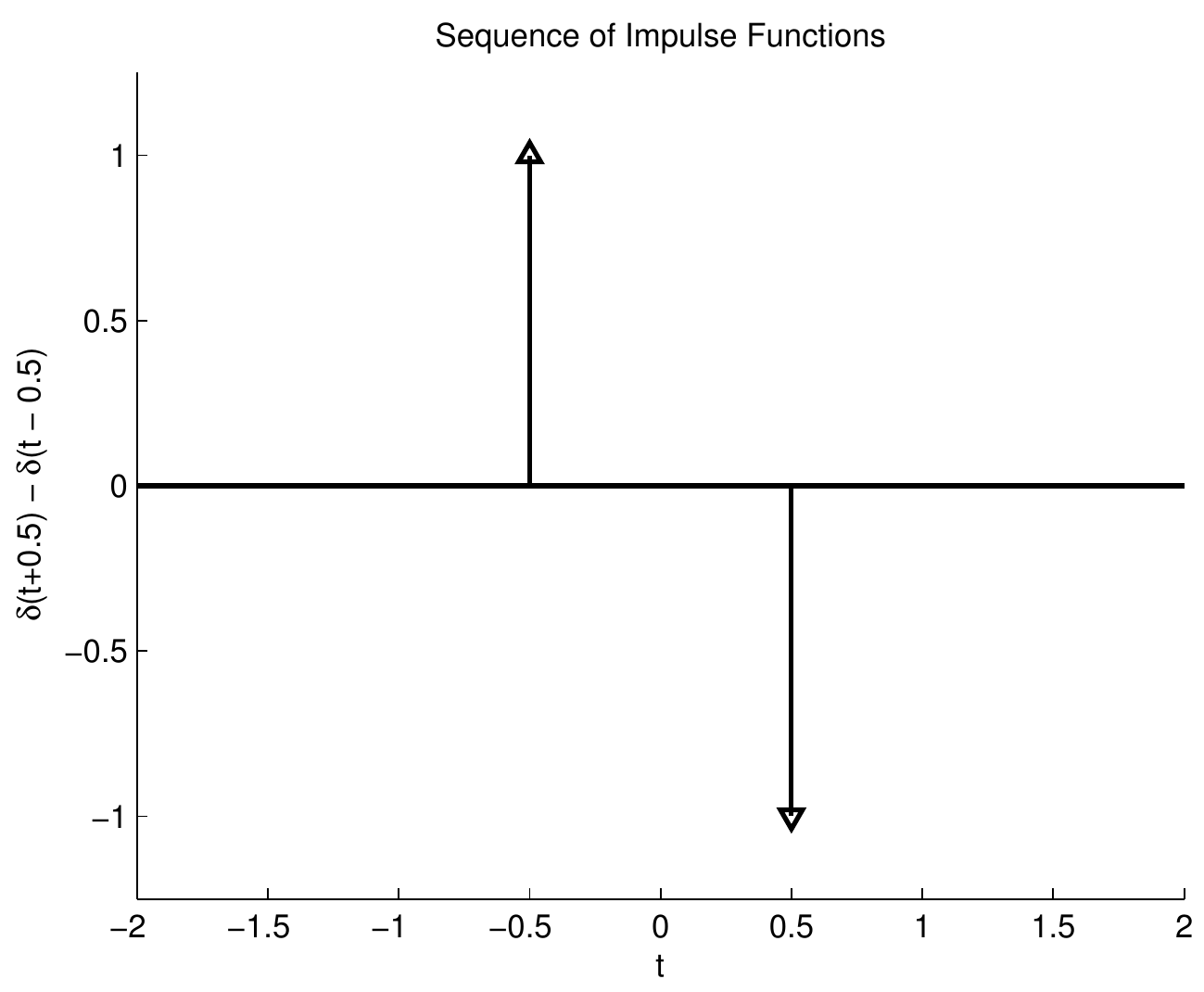}}
    \caption{Example of a control signal and its sparse spike encoding.}
  \label{fig:pulseFigure}
\end{figure}

    In a previous work, we show how square wave control signals, with abrupt switches between two states (first alluded to in \ref{enum:controlSwitchExtremes}, with an example given in figure \ref{fig:pulse}), can effectively model smooth human reaching movements \cite{Yazdani2012}. This work extends that notion by developing a method to represent neural signals via sparse usage of the Dirac delta function. These sparse signals can be thought of as encoding a series of positive (excitatory) or negative (inhibitory) neural spikes, or, more plausibly, groups of neurons spiking for brief periods. In summary, we take the square wave control signals described in \cite{Yazdani2012}, and encode them as a sparse series of Dirac delta functions, each of which signifies one of the abrupt switching points for the control signal.

    We define a \emph{sparse optimal control policy} as a control policy that meets optimality constraints with the lowest cost, as defined by the chosen cost function, that can be encoded by a finite number of discontinuous changes in the signal. An example of a signal that can be encoded as a sparse series is a rectangular pulse function, much like the control signals explored in \cite{Yazdani2012}, and shown in figure \ref{fig:pulse}. Functions like these can be encoded by impulse functions (see Figure~\ref{fig:spikedUp}). The control signals in the minimum-time control problem discussed on in section \ref{subsec:MinTimeControl} is an example of sparse control signals that are optimal in terms of state transfer time. \emph{Henseforth, we will refer to the control signal (square wave) as ``sparse", as it is easily transformed into a sparse signal consisting of Dirac delta functions.}  

%

Here we discuss sparse optimal control signals that solve the minimum-effort problem. The control signal is defined as the $n$-th order derivative in terms of position $x(t)$,

\begin{equation}\label{eq:controller}
u_{n}(t) = \frac{d^{n}}{dt^{n}}x(t).
\end{equation}
The minimum effort control problem that results in sparse control signals uses the $L_{\infty}$ norm and is written as

\begin{equation}\label{eq:sparseEffort}
\begin{aligned}
& \underset{u_{n}(t)}{\text{minimize}}
& & \sup_{0 \leq t \leq T}|u(t)| \\
& \text{subject to} & & \dot{\textbf{x}}_{n}(t) = \textbf{A}_{n} \textbf{x}_{n}(t) + \textbf{B}_{n}u_{n}(t) \\
& & & \textbf{x}_{n}(0) = \textbf{x}_{i} \\
& & & \textbf{x}_{n}(T) = \textbf{x}_{f}
 \end{aligned}
\end{equation}
where $\textbf{x}_{n}(t) = \begin{bmatrix} x(t) & \frac{d}{dt}x(t) & \frac{d^{2}}{dt^{2}}x(t) & \ldots & \frac{d^{n-1}}{dt^{n-1}}x(t) \end{bmatrix}^{T}$ is the state vector, $\textbf{x}_{i}$ and $\textbf{x}_{f}$ are the initial and final boundary conditions, $T$ is the duration of the movement (with movement starting at time $t = 0$), and $\sup_{0 \leq t \leq T}| . |$ is the $L_{\infty}$ norm. Here we consider a system that is an $n$-th order integrator, thus
\begin{equation}\label{eq:dynamicParameters}
\textbf{A}_{n} =  \begin{bmatrix}
 \textbf{0}_{(n-1)\times1} & \textbf{I}_{(n-1) \times (n-1)}\\
 0 &  \textbf{0}_{1 \times (n-1)}
 \end{bmatrix}
 \textnormal{ and } \textbf{B}_{n} =  \begin{bmatrix}  \textbf{0}_{(n-1)\times1} \\  1 \end{bmatrix}.
 \end{equation} 
 The authors of \cite{Yazdani2012} considered the special case $u_{3}(t) = \dddot{x}(t)$, ($n = 3$), and showed that this particular sparse control signal explains the trajectories of human movements better than the traditional Flash and Hogan model of equation \ref{eq:FandH}. Perhaps more importantly, sparse control signals are biologically more realistic than non-sparse signals (See the Discussion section for more on this).  As we will demonstrate, most formulations of minimum effort problems can be easily converted to generate sparse control signals, complete with the afore mentioned benefits. By simply measuring the effort term in the objective function via the $L_{\infty}$ norm (as opposed to the 2-norm), we can frame the control signal solution to these problems in terms of discontinuous switching states.

 We now show the general analytic solution for equation \ref{eq:sparseEffort}. To derive the general solution, we assume that the boundary conditions are
 \begin{equation}\label{eq:simpleBoundary}
 \textbf{x}_{i} =
  \begin{bmatrix}
  x_{i} \\
  \textbf{0}_{(n-1)\times 1}
  \end{bmatrix}
  \textnormal{ and } \textbf{x}_{f} =
  \begin{bmatrix}
  x_{f} \\
 \textbf{0}_{(n-1)\times 1}
  \end{bmatrix}.
  \end{equation}
 That is, we assume that the movement starts at rest and ends at rest. We solve the general sparse minimum effort control problem by manipulating equation \ref{eq:sparseEffort} to a form that has been previously solved. Namely, note that every optimization problem can be written equivalently as an optimization problem with a linear objective by introducing an auxiliary variable $K$ and we can equivalently express equation \ref{eq:sparseEffort} as follows:
\begin{equation}\label{eq:linearSparse}
\begin{aligned}
& \underset{u_{n}(t), K}{\text{minimize}}
& & K   \\
& \text{subject to} & & \dot{\textbf{x}}_{n}(t) = \textbf{A}_{n} \textbf{x}_{n}(t) + \textbf{B}_{n}u_{n}(t) \\
& & & \textbf{x}_{n}(0) = \textbf{x}_{i} \\
& & & \textbf{x}_{n}(T) = \textbf{x}_{f} \\
& & & | u_{n}(t) | \leq K.
 \end{aligned}
\end{equation}
where $u_{n}(t)$, $\textbf{A}_{n}$, $\textbf{B}_{n}$, $\textbf{x}_{i}$, and $\textbf{x}_{f}$ are defined as before in equations \ref{eq:controller}, \ref{eq:dynamicParameters}, and \ref{eq:simpleBoundary} respectively, and we have used the fact that $|| u_{n}(t)||_{\infty} \leq K \implies  | u_{n}(t) | \leq K$. The equivalency between equations \ref{eq:sparseEffort} and \ref{eq:linearSparse} is due to the fact that every objective can be bounded, and this bound is expressed as an additional constraint in equation \ref{eq:linearSparse}.

The optimization problem of equation \ref{eq:linearSparse} has the same form as equation \ref{eq:minimumTimeSimple}. We can therefore use the results from the minimum-time control problem and apply them here (namely that the results of Pontryagin and colleagues still hold). The difference is that in equation \ref{eq:minimumTimeSimple} the unknown is time $T$, whereas in equation \ref{eq:linearSparse} the unknown is the bound $K$ on the control signal $u_{n}(t)$. Since the dynamic system in equation \ref{eq:linearSparse} is an $n$-th order integrator, we can use the result from \cite{Feldbaum} and write the following theorem:\vspace{.2cm}\\

\newtheorem*{numSwitchings}{Number of Switches for an N-th Order Integrator. Theorem 1}
\begin{numSwitchings}
For a control problem of the type in equations \ref{eq:minimumTimeSimple} or \ref{eq:linearSparse} where the system dynamic equations are an $n$-th order integrator (as in equation \ref{eq:dynamicParameters}), then the number of switchings in the control signal is exactly $n+1$ and the control signal is symmetric.\vspace{.2cm}\\
\end{numSwitchings}

\begin{figure}
  \centering
  \subfigure{\label{subfig:jerk}\includegraphics[scale = .5]{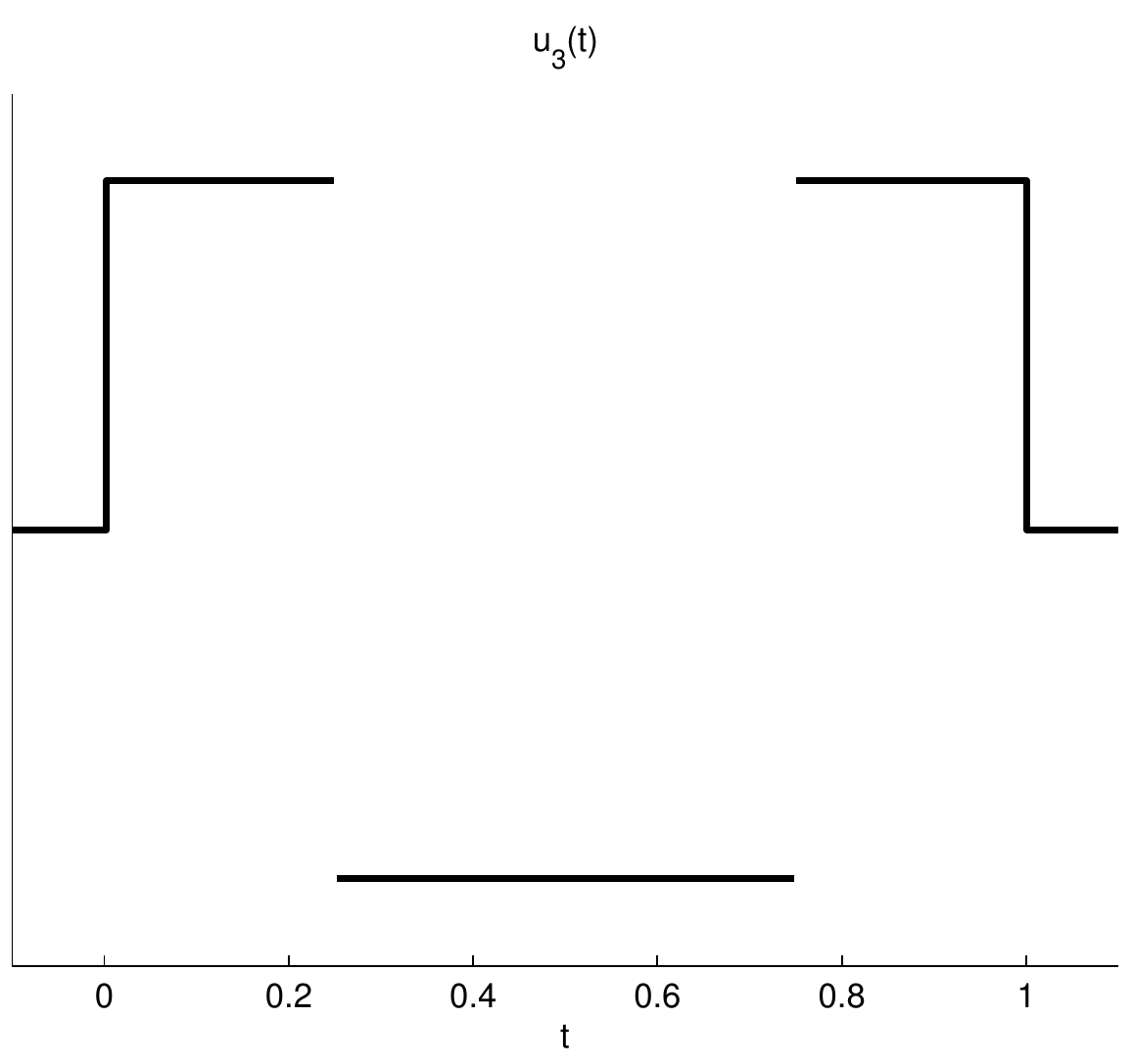}}        \hspace{15mm}
  \subfigure{\includegraphics[scale = .5]{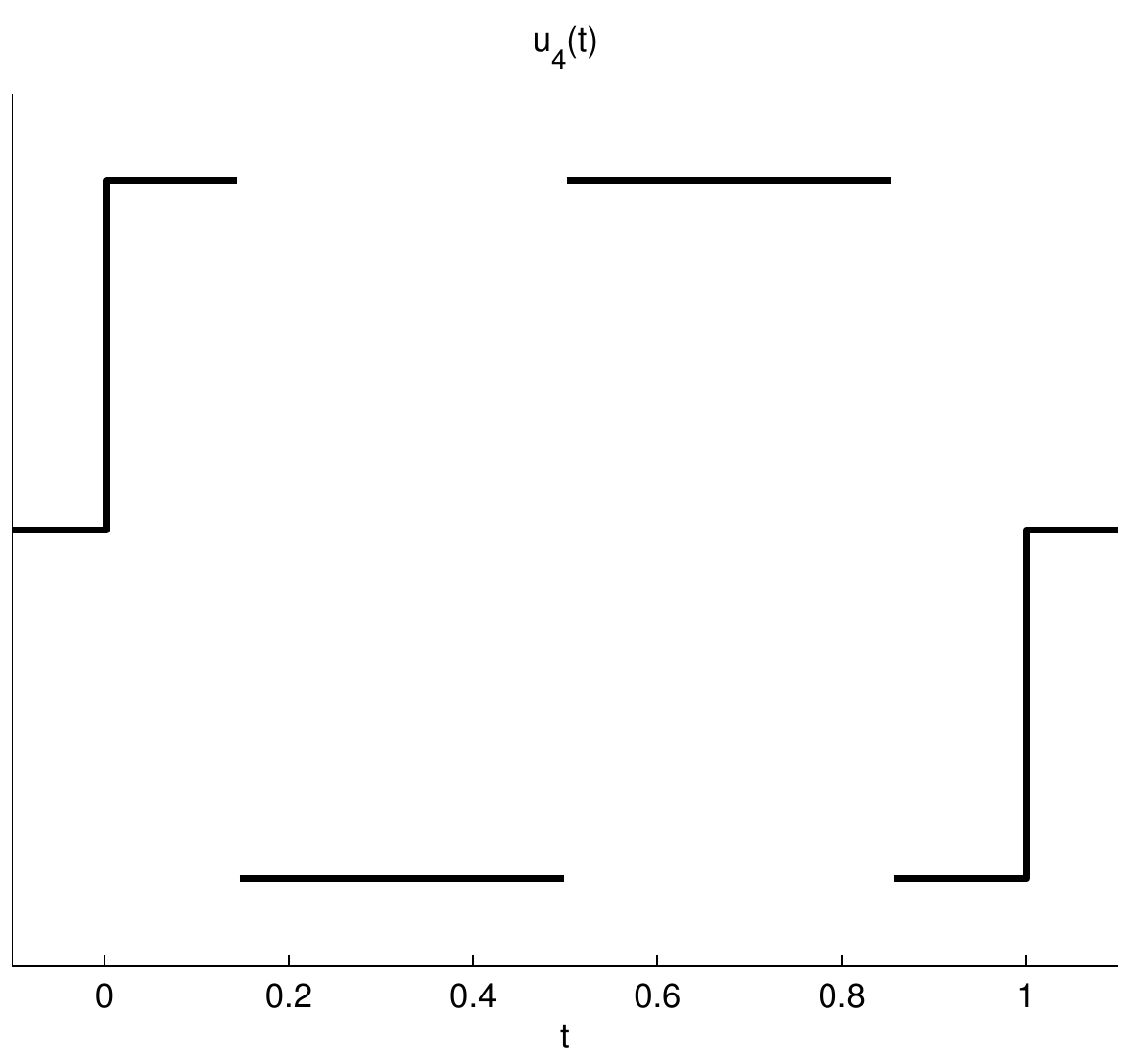}}
  \subfigure{\label{subfig:crackle}\includegraphics[scale = .5]{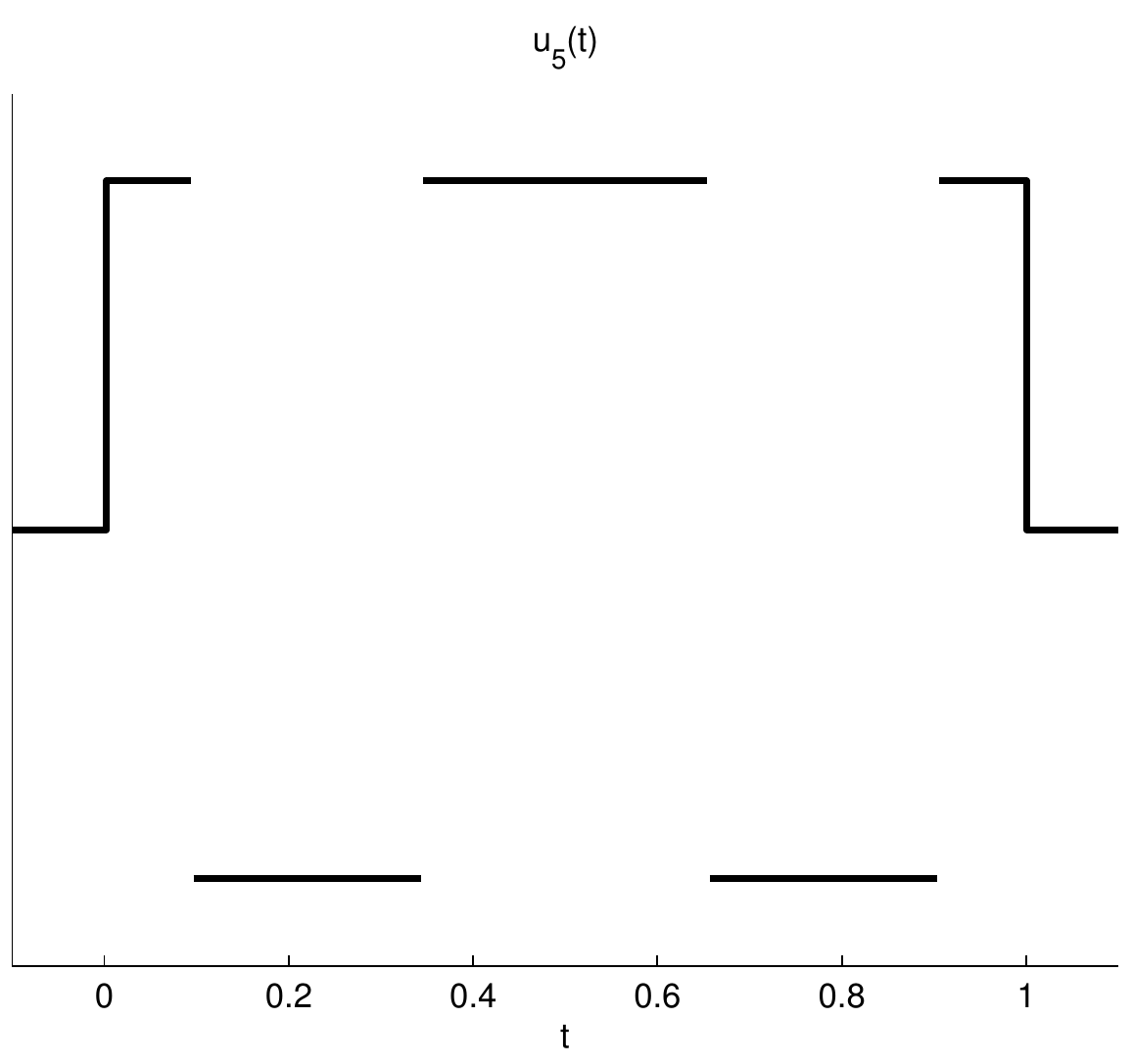}}  \hspace{15mm}
  \subfigure{\includegraphics[scale = .5]{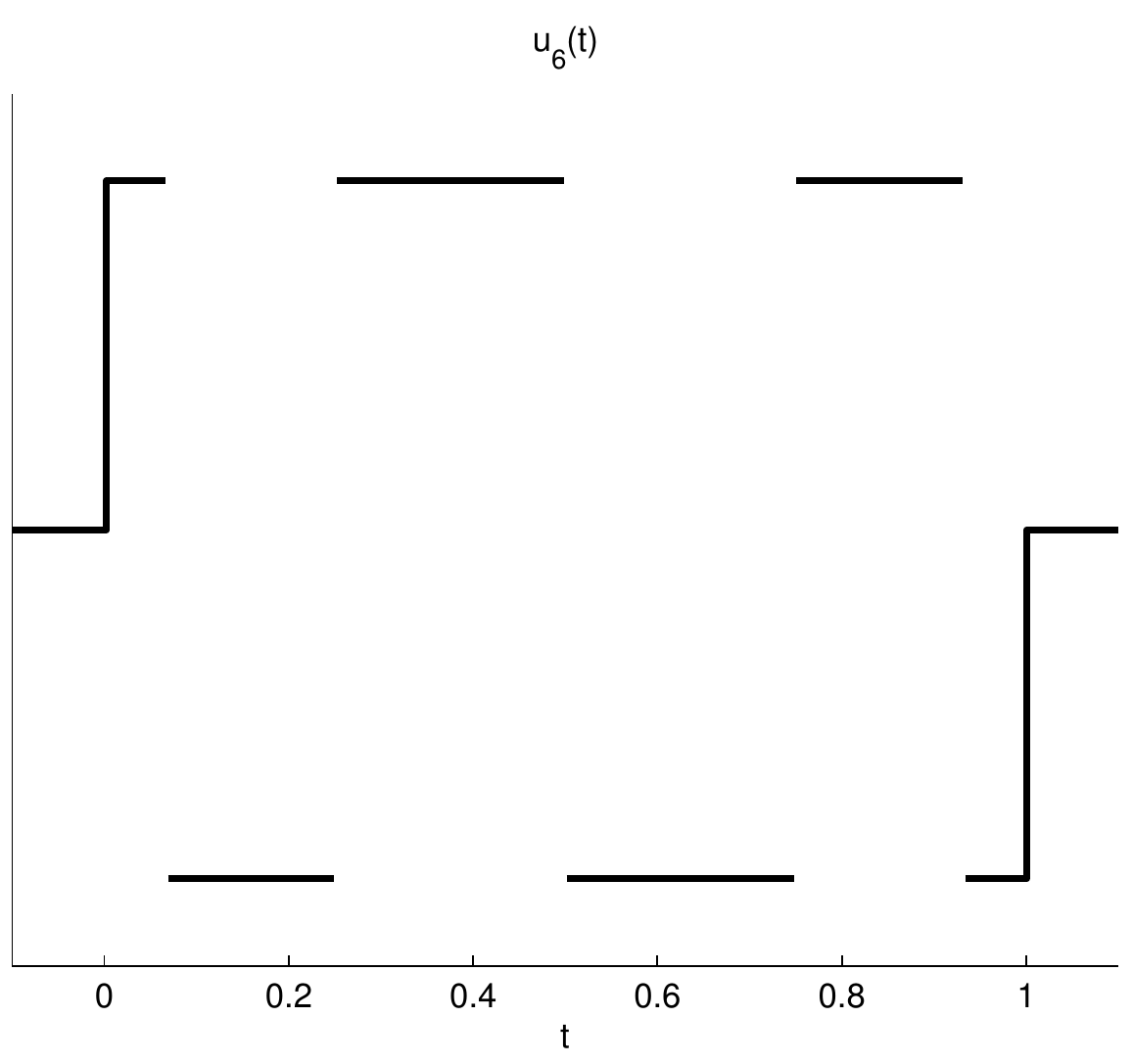}}
    \caption{Examples of sparse optimal control signals $u_{i}(t)$ for $i = 3, \ldots, 6$. Shown are movements that start from $t = 0$ and end at $t = 1$. }
  \label{fig:sparseControlSignals}
\end{figure}

In other words, as initially discussed in \ref{enum:controlSwitchesLimited}, as the order of the control signal increases (as $n$ increases), the number of switches in the control signal increases by the same amount. \cite{Svinin} solved the general $n$-th order minimum-time control problem of equation \ref{eq:minimumTimeSimple} for an $n$-th order integrator. We adapt their results for the general $n$-th order minimum-effort control problem of equation \ref{eq:linearSparse} and summarize the solution as follows:
\begin{equation}\label{eq:sparseAmplitude}
\boxed{K^{*}_{n} = \frac{2^{2(n-1)}(n-1)!(x_{f} - x_{i})}{T^{n}}}
\end{equation}
\begin{equation}\label{eq:sparseSwitch}
\boxed{t_{i}^{*} = T\text{sin}^{2}\left( \frac{\pi i}{2n} \right), \quad i = 0, \ldots, n}
\end{equation}
where $K^{*}$ denotes the optimal amplitude or bound on the control signal, and $t_{i}^{*}$ denotes the optimal switching times. Figure \ref{fig:sparseControlSignals} shows examples of several sparse optimal control signals.


%
%
%

\subsection{Sparse Optimal Control Signals in Fast Human Movements}
The sparse optimal control signals introduced in section \ref{sec:SparseControlPointToPoint} are not only optimal with respect to a minimum-effort objective, but are also more biologically plausible when compared with non-sparse signals. Sparse optimal control signals can be are those that can be efficiently represented with Dirac delta functions, which resemble neuronal bursts or spikes. If we treat each spike as an idealized Dirac delta function, as done in \cite{Dayan}, and as visualized in Figure~\ref{fig:spikedUp}, then a spike sequence that represents an $n$-th order optimal control signal can be expressed as
\begin{equation}\label{eq:spikeTrain}
\rho_{n}(t) = K_{n}^{*}\sum_{i = 0}^{n} (-1)^{i}\delta(t - t_{i}^{*})
\end{equation}
where $K_{n}^{*}$ and $t_{i}^{*}$ can be found from equations \ref{eq:sparseAmplitude} and \ref{eq:sparseSwitch} respectively. The spike train represented by equation \ref{eq:spikeTrain} is not postulated to be from a single neuron, but rather a population of excitatory and inhibitory neurons forming a network.

The work of \cite{Ben-Itzhak, Yazdani2012} showed that a sparse optimal control signal that corresponds to jerk (expressed as $\rho_{3}(t)$ in the notation of equation \ref{eq:spikeTrain}) can model fast human movements with greater accuracy than the smooth control signal that results from using the $L_{2}$ norm. We now also propose that the control signals the nervous system uses are not limited to the jerk control signal. There is nothing preventing the nervous system from using a higher-order control signal (see Figure~\ref{fig:resultsSnapCrackle} for comparisons higher-order derivative control signals). With each increase in the order of the control signal, the number of spikes increases and the timing of those spikes changes (as shown in Figure \ref{fig:sparseControlSignals}). Because each wave form is different, these high-order derivative control signals can form a basis set, the elements of which are combined to form a subspace of control signals. The neuroscience motor control literature commonly refers to the elements in such basis sets as ``motor primitives". These primitives are combined to control a variety of animal movements\cite{GiszterNeuro}.

\subsection{Application of Sparsity to Extensions of Minimum Effort Control}
The minimum effort control problem proposed in section \ref{subsec:miniumEffortControl} can be extended in numerous ways, e.g. \cite{KawatoMinTorqueChange1989,BigDaddyKMTH, Yazdani2012, Ben-Itzhak, EmkenGreedyOptimizationErrorEffort,Flash,Liu2007}. These works and others, account for various aspects of movement and draw different conclusions regarding the nature of the motor system. For example, in \cite{Liu2007}, several types of reaching movements under various conditions are analyzed and modeled via an extension of the minimum effort problem. The types of movements included both two and three dimensional reaching, with and without target perturbation, with and without obstacle avoidance, and under various instructions to the subject regarding how the target should be impacted. The following is a simplified version of the model in \cite{Liu2007} that maintains its core concepts: a term for effort and a term for final state error. Equations \ref{eq:LiuTodorovSimplification} and \ref{eq:LiuTodorovSparse} give an example of how minimum effort control problems can easily be adapted to our method of generating sparse signals.

\begin{equation}
\label{eq:LiuTodorovSimplification}
\begin{aligned}
& \underset{u(t)}{\text{minimize}}
& & ||\textbf{x}(T) - \textbf{x}_{f}||_{2}^{2} + w_{\text{effort}}\int_{0}^{T}u(t)^{2}dt   \\
& \text{subject to} & & \dot{\textbf{x}}(t) = A \textbf{x}(t) + Bu(t) \\
& & & \textbf{x}(0) = \textbf{x}_{i}
 \end{aligned}
\end{equation}
Equation \ref{eq:LiuTodorovSimplification} is the same as the minimum effort control problem discussed earlier, with the exception that hard equality constraints (the end-point boundary conditions) are now ``soft" constraints and are penalized as a cost. The $w_{\text{effort}}$ term dictates a trade off between minimizing effort and meeting the final boundary conditions. To have a sparse implementation of the above, we use the infinity-norm (sup) as before:

\begin{equation}
\label{eq:LiuTodorovSparse}
\begin{aligned}
& \underset{u(t)}{\text{minimize}}
& & ||\textbf{x}(T) - \textbf{x}_{f}||_{2}^{2} + w_{\text{effort}}\sup_{0 \leq t \leq T}|u(t)|  \\
& \text{subject to} & & \dot{\textbf{x}}(t) = A \textbf{x}(t) + Bu(t) \\
& & & \textbf{x}(0) = \textbf{x}_{i}
 \end{aligned}
\end{equation}
This optimization problem, similar to the sparse minimum effort problem above in equation \ref{eq:sparseEffort} can be written as follows:

\begin{equation}
\begin{aligned}
& \underset{u(t), K, K_{1}, K_{2}}{\text{minimize}}
& &  K  \\
& \text{subject to} & & \dot{\textbf{x}}(t) = A \textbf{x}(t) + Bu(t) \\
& & & \textbf{x}(0) = \textbf{x}_{i} \\
& & & K = K_{1} + K_{2} \\
& & & ||\textbf{x}(T) - \textbf{x}_{f}||_{2}^{2} \leq K_{1} \\
& & & |u(t)| \leq \nicefrac{K_{2}}{w_{\text{effort}}} \\
& & & K_{1} \geq 0, K_{2} \geq 0
 \end{aligned}
\end{equation}
As before, we are using the property that every optimization function can be written equivalently as an optimization problem with a linear objective by introducing an auxiliary variable. Again, we have used the property that $w_{\text{effort}}\sup_{0 \leq t \leq T}|u(t)| \leq K_{2} \Rightarrow w_{\text{effort}}|u(t)| \leq K_{2}$. Therefore, we also have a ``bang-bang" solution, since the control signal is hard bounded.

\section{Materials and Methods}
The human arm movement data for this work was originally collected by Karniel and Mussa-Ivaldi and used in their 2002 paper to investigate the nervous system's ability to adapt to perturbations. We used a subset of this data that was relevant for our study of fast movements (the baseline unperturbed movements) and summarize their experimental setup below and refer the reader to \cite{BigDaddyKandMussa2002} for a more complete description.

\begin{figure}
  \centering
  \subfigure[Experimental setup for collection of fast reaching movement data. Subjects sat down and held a manipulandum with their hands which they could maneuver about a 2D plane positioned in front of them perpendicular to their torsos. Regions 'A', 'B' and 'C' (Figure adapted from \cite{Yazdani2012} with permission.)]{\label{fig:manipSetup}\includegraphics[width=0.3\textwidth]{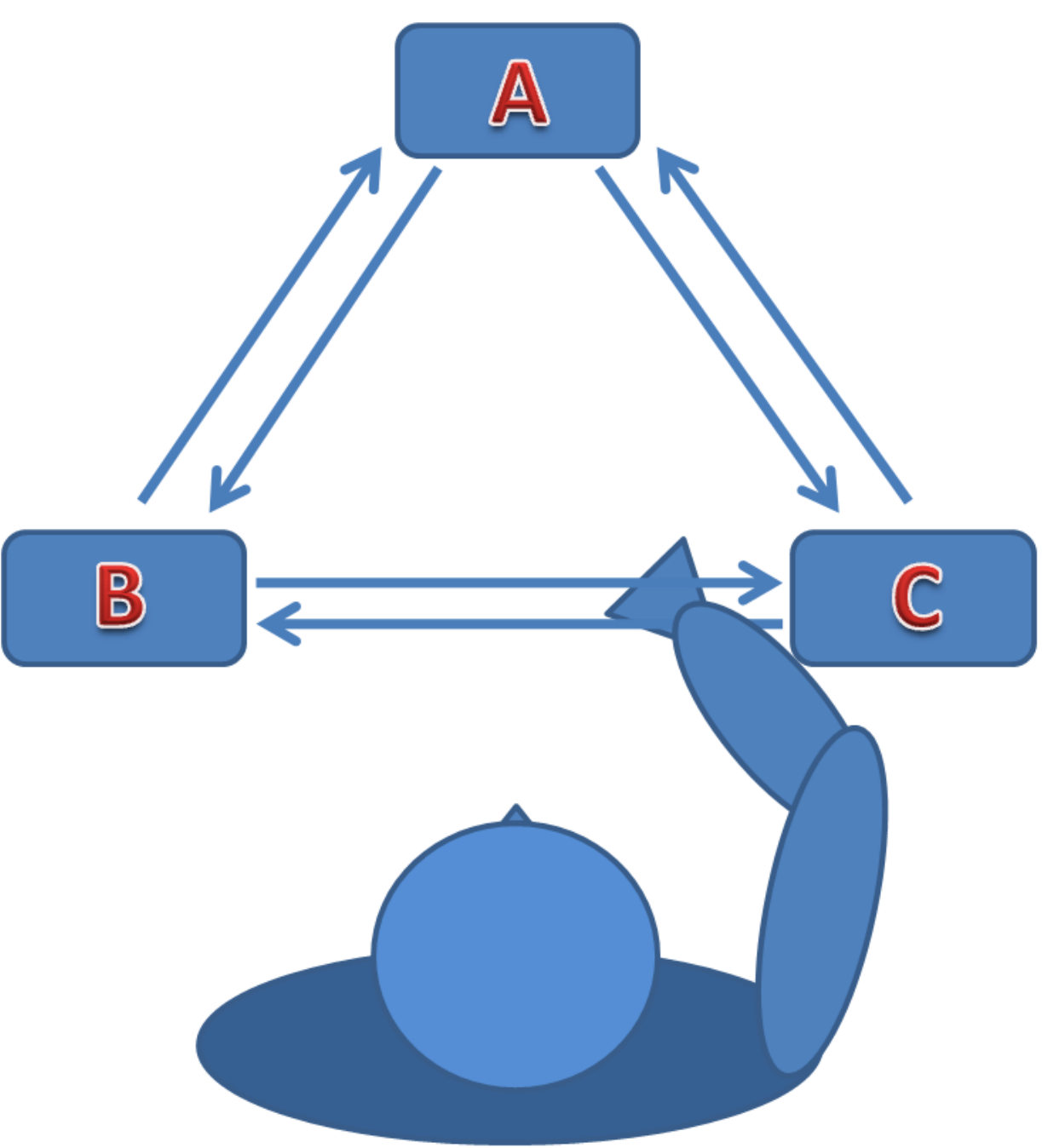}}
  \hspace{5mm}
  \subfigure[A velocity profile for a typical hand movement trial. We attempt to identify the ``ballistic" (fast) portion of the movement by using an onset and offset detection algorithm to automatically detect the beginning and end of the fast reaching portion of the movement. The ``onset" is indicated by the leftmost red line and the ``offset" is indicated by the rightmost red line. The ``corrective region" refers to the time during which the subject attempts to correct any over or undershoot of the target, this region is not modeled. Data between the red lines is modeled, the rest is discarded.]{\label{fig:typicalTrial}\includegraphics[width=0.5\textwidth]{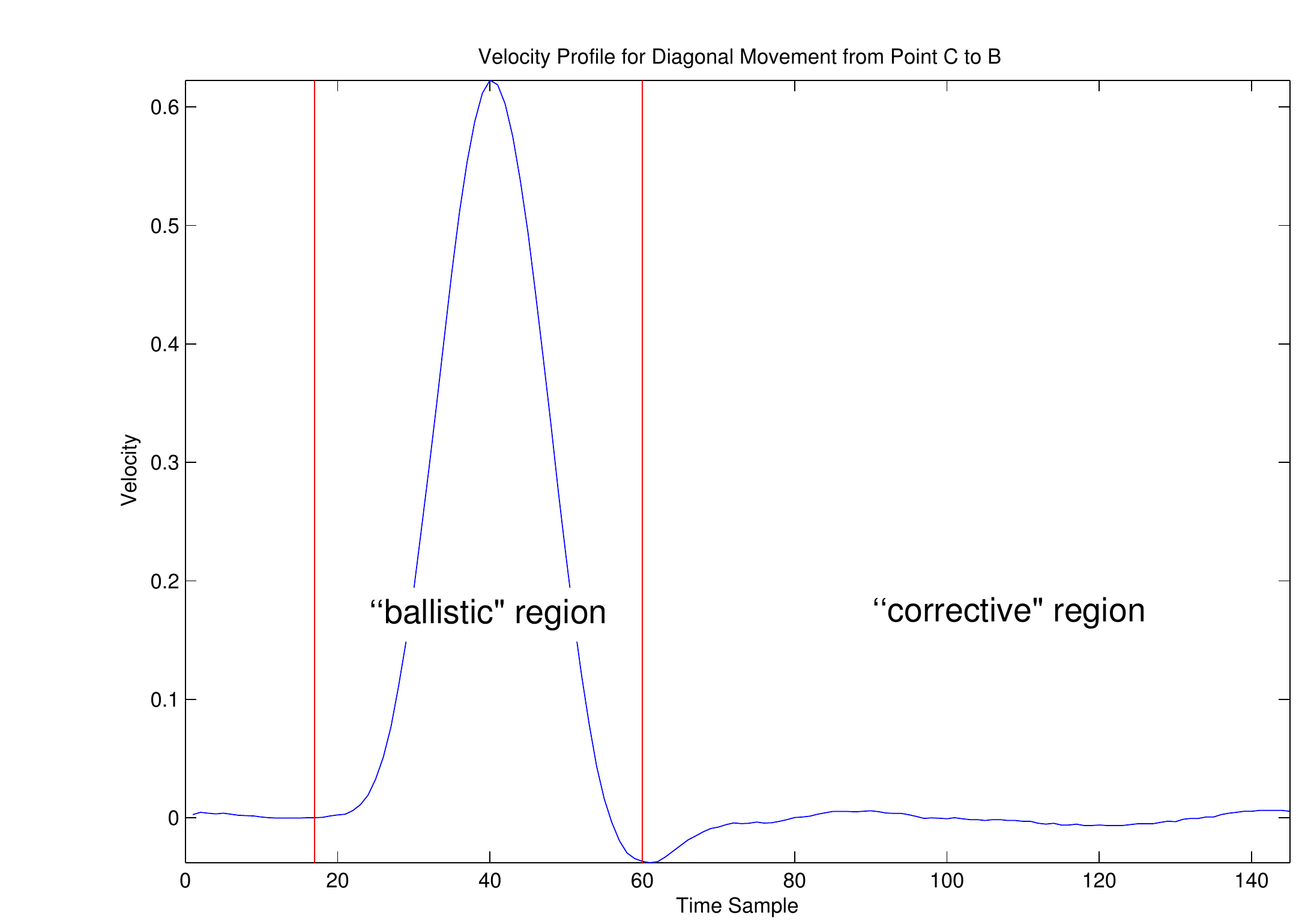}}
    \caption{Setup for the recording of human reaching movements and a typical velocity profile from those data for one reaching movement trial.}
  \label{fig:animals}
\end{figure}

Five subjects participated in an experiment involving a manipulandum that restricted their movements to a the horizontal plane in front of the seated subjects (subjects participated separately in these experiments). During each trial, the subject watched a screen that displayed the position of their hand and the manipulandum in relation to three positional markers A, B and C. Each marker was separated by 10 cm and formed an equilateral triangle (see Figure~\ref{fig:manipSetup}). For each trial, the subject was instructed to move the on-screen representation of the manipulandum from one target to another in about one third of a second with a tolerance of $\pm$50ms. At the end of each trial feedback was given indicating if the subject had reached the target and also if the execution of their movement was within the allowed time window. The trajectories were recorded for all six possible movement types for all subjects over the course of four days. In all, there are 366 trials for the 5 subjects.

%

For each trial, we only select the so-called ``ballistic" portion of the movement (see Figure~\ref{fig:typicalTrial}). That is, we select the portion of the trial where movement had started and the movement had completed its feedforward portion, the portion of the movement that was preplanned and not affected by peripheral feedback. Part of the justification for this understanding of fast reaching movements is that they are happening too quickly for a corrective proprioceptive signal to make a meaningful difference, as discussed in \cite{gerdes1994noFeedback , keele1968noFeedback}. Thus we do not model the ``corrective" portion of the movement that likely involves additional feedback information from the subjects' visual system and limbs in order to fix any error when attempting to reach to the targets. For this reason, we frame our approach as a feed forward control problem, because feedback is not involved in the movements. This is the same approach used in several studies including \cite{Yazdani2012,karniel1997FeedforwardModel,Ben-Itzhak}.

There are various methods for movement onset and offset detection \cite{BotzerKarniel, StaudeOnsetDetectionHumanMotor, StaudeOnsetDetectionVoluntaryMotorResonse}, and there is no standard technique for choosing the relevant portion of a movement since the definition of what is relevant may change from study to study or from one movement type to another.

We approach finding the start and end of movement by finding the point in time when the velocity profile has reached it's peak velocity. Fast movements always have a unique global maximum in the trial (unless the trial is an outlier) so finding the time at which this maximum occurs is unique. Once this point in time is found, we proceed to consider velocity samples before and after the peak velocity and find the sample that falls below a pre-determined threshold. This methods extracts the ballistic region of the movement.

The optimization procedures were implemented using CVX and Matlab. CVX is package for specifying and solving convex optimization problems \cite{grant2011cvx}.


\section{Comparison of Sparse and Non-Sparse Model Predictions to Human Reaching Movements}
\begin{figure}[!htb]
\centering
\includegraphics[scale=0.35, angle = 270]{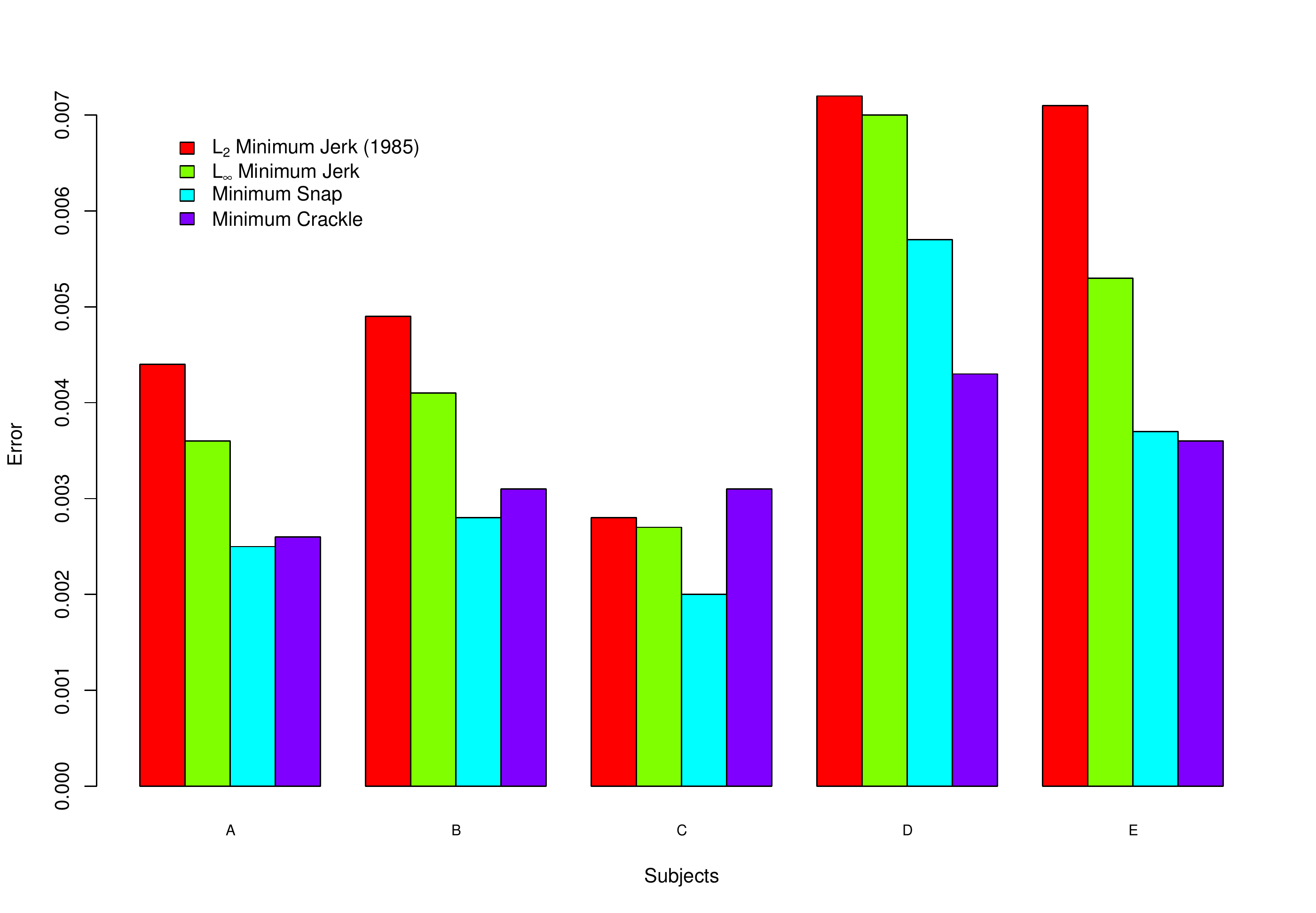}
\caption{Comparison of average mean squared error between reaching trials for all five subjects and predictions made on those trials by four models. \textcolor{red}{Red} bars indicate MSE for the $L_{2}$ norm based \textcolor{red}{minimum jerk model} (\emph{continuous}). MSEs shown in \textcolor{green}{green}, \textcolor{cyan}{cyan}, and \textcolor{purple}{purple} are all based on the $L_{\infty}$ norm, and thus result from \emph{sparse} control signals. They represent  \textcolor{green}{minimum jerk} (\emph{sparse}), \textcolor{cyan}{minimum snap}, and \textcolor{purple}{minimum crackle}, respectively.
The sparse control signals on average have smaller error than the continuous control signal (resulting from use of the $L_{2}$ norm). In all comparisons but one, the sparse control signals result in lower MSE than the continuous control signal. The one exception is in the case of subject C where we see the the minimum crackle sparse control signal produce a greater error that the $L_{2}$ norm minimum jerk (continuous) control signal, however, the difference between the errors is not statistically significant (as determined by a Wilcoxon rank-sum test).}
\label{fig:resultsSnapCrackle}
\end{figure}

Figure~\ref{fig:resultsSnapCrackle} shows the average mean squared error (MSE) between the human subjects' trial velocity profiles and the velocity profiles generated by four computational models. We computed the MSE between the models' velocity profiles and those of the human subjects across time steps of the recorded movement. For all models, portions of the velocity profile which were forced to be accurate due to setting boundary conditions were not included in the MSE calculation. The figure highlights the canonical minimum jerk (as measured by the $L_{2}$ norm) model which results in a continuous, non-sparse control signal. We compare this model with one that minimizes jerk as measured by the $L_{\infty}$ norm. In all cases the sparse signal generated by this model has a lower error than the minimum jerk $L_{2}$ norm model. This is not to say that the $L_{\infty}$ norm model is decidedly better in all cases, but it is in most, and at the very least it performs similarly to the $L_{2}$ norm model while retaining the characteristic of an intuitive mapping to a spike train representation.

We also contrast the results of the $L_{2}$ norm model with three additional models that minimize snap, crackle, and pop (fourth, fifth, and sixth derivatives of position) as measured by the $L_{\infty}$ norm (resulting in sparse control signals). Errors of all models employing sparse control signals are smaller than those generated by the minimum jerk model employing a non-sparse control signal in all cases but one. In this case, for subject C, the sparse crackle ($L_{\infty}$) based model has a higher error than the non-sparse jerk ($L_{2}$ norm) based model, however, the difference between the non-sparse minimum jerk error and sparse crackle error is not statistically significant (by a Wilcoxon rank-sum test). This demonstrates that regardless of the chosen derivative of position used as the control signal (jerk, snap, or crackle), sparse signals are effective control strategies.


\section{Discussion}

\subsection{Sparse Signals and Their Biological Plausibility}
Converting traditional minimum effort models to their sparse counterparts via methods outlined in this work can bring these models closer to a plausible biological interpretation in several ways. At the level of the observation of human movement, studies have indicated that human subjects use sparse (intermittent) control strategies for ballistic movements to control activities that are continuous in nature \cite{LoramIntermittantControlPhysiological, Ben-Itzhak}.

At the neural system level, there is evidence that various neural structures exhibit intermittent behavior. For example, the basal ganglia have been shown to be key components in the control of movement. Inputs to the basal ganglia, arriving from a large portion of the cerebral cortex, exhibit intermittent behavior \cite{redgrave1999basalSwitching}. More generally, the basal ganglia as a whole are thought by most to be a network that switches between well defined states, in an intermittent fashion \cite{redgrave1999basalSwitching}. Another example that is directly applicable to this work examines a primate tasked with making reaching movements towards two \emph{possible} targets that the animal is accustomed to. For a given trial, the ``correct" target is not initially known to the animal. At this point, two sustained signals (sustained neural activity), representing each potential target, are present in its pre-motor cortex. Once the ``correct" target (for a given trial) is revealed, an abrupt switch occurs where the neural signal representing the ``incorrect target" is suppressed and the ``correct" neural signal remains \cite{cisek2005neuralMultiSignalSwitch}. This is is clearly an abrupt switch between two states, as suggested by the sparse model of reaching tasks outlined in this work.

\begin{figure}
  \centering
  \includegraphics[width=0.5\textwidth]{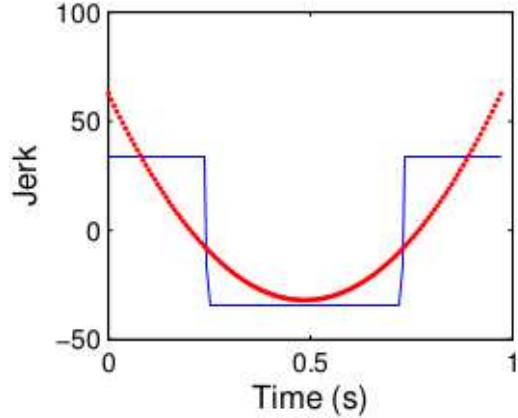}
  \caption{The \textcolor{red}{red} plot shows a minimized jerk control signal as measured by the $L_{2}$ norm. Note that it is parabolic and continuous (non-sparse). The \textcolor{blue}{blue} plot shows a minimized jerk control signal as measured by the $L_{\infty}$ norm. Note the distinct qualitative difference between the two. The $L_{\infty}$ based signal is encoded by a finite number of abrupt changes, which can be encoded . Such a signal can be characterized by Dirac delta functions as show in Figures \ref{fig:pulse} and \ref{fig:spikedUp}. These spikes resemble neuronal spikes. In contrast, the $L_{2}$ norm based signal and has no obvious mapping to the physiology of neurons. (figure originally used in \cite{Yazdani2012})}
  \label{fig:2NormVsInfinityControlSig}
\end{figure}

At the neuronal level, it has been shown in \cite{LoewensteinBistabilityPKJCells} that Purkinje cells in the cerebellum (well known to be involved with motor control) exhibit bistability. That is, they have two modes of operation, each of which persist until a switching event occurs. This event consists of \emph{brief neural pulses}, which switch the Purkinje cell back and forth from a highly active state to an inactive state. The modeled spike trains suggested in this work (e.g. Figure \ref{fig:spikedUp}) may be interpreted as single neurons or they may be interpreted as ensembles of neurons working in concert. In either case, the the pulses controlling the bistable state of the Purkinje cell can be represented by Dirac delta functions, which is an accepted technique to mathematically represent spike trains \cite{Dayan}. In this way, these signals more closely mirror the physiology of neurons when compared to their non-sparse counterparts (as shown in Figure \ref{fig:2NormVsInfinityControlSig}). Conceptually, there is a mapping from sparse switches (Dirac deltas) to neuronal spikes, or groups of neurons spiking. Furthermore, spike timing, and its relevance in the neural coding of control information is directly represented, which is not the case with the continuous control signal method. In addition, the sparsity of these signals simplify the necessary output of a neural circuit used to drive motor function. For example, it would require only four spikes to encode the signal shown in Figure \ref{subfig:jerk}. This concept lends itself to a hierarchical control structure employing ``higher level" neural motor control structures that focus on learning and producing simple switching times, which drive and offload more complicated tasks and signal processing to lower level structures (as discussed in \cite{BigDaddyKMTH}) which exist in the brain stem and spinal cord. To reiterate, we are not stating that the models in this work imply that a single neuron is driving any kind of motor function, or that these signals \emph{directly} drive muscles. Rather, that, the concepts here are an abstraction which indicates that at some level of a hierarchy of neural motor control hierarchy, this sparse control approach is plausible.

It is clear that the timing of spike trains plays an important role in their meaning and information content \cite{DiLorenzoSpikeTiming2013, GerstnerMarkramFiringRates}. Unlike continuous control signals, the sparse signals discussed here have very specific switching times, each switching time being integral to the character of the resulting modeled movement. Because of this, the sparse approach can model spike timing as it pertains to neural coding of information, while the continuous models have no explicit representation of spike timing or abrupt system level switches.

It has been proposed that signal dependent noise contributes to the variability of observed movements \cite{Harris1998}. Both \cite{JonesWolpertSigDependNoiseForceProduction, WolpertScalingMotorNoiseMuscleStrength} provide empirical support for this proposal. This viewpoint correlates neural signal magnitude with the level of noise in the system, and therefore the accuracy of the movement. In other words, signals that are extremely strong (i.e. all available neurons for a given task are firing), inherently create more noise in their own system, lessening the accuracy of the movement. Using the $L_{\infty}$ norm to generate sparse control signals has the additional benefit of setting a upper bound on the magnitude of the control signal, thus upper bounding the signal-dependent noise, and increasing accuracy. Thinking of the two ideas (upper bounding a control signal, and signal dependent noise) in this way may reconcile any perceived discrepancy between them. Figure \ref{fig:2NormVsInfinityControlSig} shows the continuous signal exceeding the absolute magnitude of the sparse signal in two locations, thereby creating more noise in the system at those points. There is nothing explicitly preventing the continuous signal from becoming arbitrarily large at any point along the way, allowing the noise associated with large signals to reach arbitrarily large levels. Using $L_{\infty}$ norm avoids this problem, as it sets a cap on the absolute magnitude of the control signal, and therefore, the signal dependent noise. In addition, preventing arbitrarily large signal magnitudes is in line with the physiology of (populations of) neurons, which can only fire with a maximum strength and frequency.

\subsection{Future Work}
Future work should consider extensions of the simple minimum effort cost functions used in this work in order to describe a richer set of movements. This can be accomplished by simply penalizing the norm of the effort term in any cost function (such as those outlined in \cite{Berret}) with an $L_{\infty}$ norm instead of the $L_{2}$ norm that is typically used. Additionally, the possibility be should explored that some type of \emph{combination} of signals of the type shown in Figure \ref{fig:sparseControlSignals} may be advantageous when modeling a richer set of movements. Combining select signals might form a basis set which would provide a larger subspace of signals that, by extension, controls a larger subspace of human movements than have been explored in this work. We envision a set of sparse signals generated by separate populations of neurons which produce sparse control signals that are combined (linearly or otherwise). This ``combination of signals" concept is inline with the idea that the CNS achieves control through relative activation of motor primitives (as described by Giszter \cite{GiszterNeuro}). He empirically illustrates how spinal motor primitives can be thought of as a basis set that can be combined in varying degrees to achieve a desired movement. However, in addition to controlling the relative magnitude of activation between primitives, changing the \emph{timing} of the activation of the primitives is equally, if not more important, as only a proper sequence of motor primitive activation will provide the desired motor output. A basis set of signals as described in this work would be advantageous as it would require a minimum amount of neural structures to control human movement. This combination of ``on-off" sparse signals might be thought of as controlling motor primitives by varying the activation times between them. Future work should consider combining such sparse signals to describe a richer set of movements. Furthermore, since this work is an abstraction of neural control signals, and it models reaching movements well, the abstraction may be a useful platform to apply to robotic motor control.


\section{Acknowledgments}
We are grateful to Amir Karniel for sharing the reaching movement data used in this work. We thank Robert Hecht-Nielsen for his guidance and Dr. William Lennon for useful discussion. We are especially grateful to Dr. Thomas McKenna and the Office of Naval Research for their support of this work.

\bibliographystyle{apalike}

\bibliography{sparseOptimalControlDissertation}



\end{document}